\documentclass[twocolumn,showpacs,preprintnumbers,amsmath,amssymb]{revtex4}

\usepackage{epsfig}
\usepackage{graphicx}
\usepackage{dcolumn}
\usepackage{bm}
\usepackage{subeqnarray}
\usepackage{subeqn}
\begin{document}


\title{Test of Fermi gas model and plane-wave impulse approximation \\
against electron-nucleus scattering data}

\author{A.V. Butkevich} 
   \email{butkevic@inr.ru}
\author{S.P. Mikheyev}
    \email{mikheyev@pcbai10.inr.ruhep.ru}
\date{\today}
\affiliation{ Institute for Nuclear Research, 
Russian Academy of Science, 
60th October Anniversary p.7A,
Moscow 117312, Russia}

\begin{abstract}
A widely used relativistic Fermi gas model and plane-wave impulse 
approximation approach are tested against electron-nucleus scattering data. 
Inclusive quasi-elastic cross section are calculated and compared with 
high-precision data for $^{12}$C, $^{16}$O, and $^{40}$Ca. A dependence of 
agreement between calculated cross section and data on a momentum transfer 
is shown. Results for the $^{12}$C($\nu_ {\mu}, \mu ^-$) reaction are 
presented and compared with experimental data of the LSND collaboration.   
\end{abstract}

\maketitle

\section{Introduction}

A precise description of neutrino-nucleus ($\nu A$) interactions in 
intermediate energy region is important for the interpretation of 
measurements by many neutrino experiments. The understanding of their 
sensitivity to neutrino properties, evaluation of the neutrino fluxes and 
spectra depend on the accuracy to which the $\nu A$ cross sections are known. 
This is in particular crucial in analysis of the long-base line neutrino 
oscillation 
experiments in which the parameter of neutrino oscillation $\Delta m^2$ is 
determined using the total number of detected events and the distortions 
in the energy distribution of the detected muons caused by neutrino 
oscillation. 
On the other hand the neutrino-nucleus cross sections contain contributions 
from both axial-vector and vector currents and thus provide 
complementary information to that provided by electron-nucleus scattering, 
which is sensitive only to the nuclear vector current.

In many experiments the neutrino fluxes in sub-GeV and GeV energy region are 
used. At such energies the charged-current quasi-elastic (QE) neutrino-nucleus 
interactions give the main contribution to the detected events. Sizable 
nuclear effects have been observed in lepton scattering off nucleus at 
energies less than a few GeV. They indicate that the nuclear environment plays 
an important role even at energies and momenta larger than those involved in 
typical nuclear ground state processes. The understanding of nuclear 
effects is relevant for the long-base line neutrino experiments in order to 
control the corresponding systematic uncertainties.

Many Monte-Carlo (MC)~\cite{REV1} codes developed for simulation of the  
neutrino detectors response are based on a simple picture of a nucleus as a 
system of quasi-free nucleons, {\it i.e.} a relativistic Fermi gas 
model (RFGM) 
~\cite{REV2}. It takes into account Fermi motion of nucleons inside the 
nucleus and Pauli blocking effect. Unfortunately the uncertainties of
a few $\nu $A scattering data at low and intermediate energies don't allow us 
to estimate the accuracy of this model. On the other hand, as follows from 
vast high-precision electron scattering data the RFGM neglects some 
important nuclear effects. So, the model which is used for calculation of 
neutrino scattering off 
nucleus should first be tested against electron scattering data.

In the present work we intent to evaluate the accuracy of the Fermi gas 
model and plane-wave impulse approximation (PWIA) [3,4,5]. 
The electron QE cross sections are calculated in the 
framework of these models and compared with high-precision  
data for different nuclei. This comparison shows that the agreement
between predictions of these models and data depends significantly 
on the momentum transfer to the target.
We applied the RFGM and plane-wave impulse approximation to 
$^{12}$C($\nu_{\mu},\mu^-$) reaction also. 

The formalism of an inclusive charged current lepton-nucleus QE scattering is 
given in Sec.2. Results are presented and discussed in Sec.3 and some 
conclusions are drawn in Sec.4.
      
\section{Formalism of the inclusive quasi-elastic scattering}      

In electromagnetic and weak charge current process electrons (neutrinos) 
interact with nuclei via 
the exchange of photons or W-boson and charged leptons are produced in the 
final state. In an inclusive reaction, in which incident 
electron ($\sigma ^{el}$) or neutrino ($\sigma ^{cc}$) with four-momentum 
$k_i=(\varepsilon _i,{\bf k}_i)$ is absorbed by nucleus with mass $m_A$ 
and only the out-going lepton with four-momentum 
$k_f=(\varepsilon _f,{\bf k}_f)$ and mass $m_l$ is detected, the cross 
section is given by contracting the lepton tensor and the nuclear 
tensor
\begin{equation}
d\sigma^{el}=\frac{\alpha^2}{Q^4}\frac{1}
{\vert {\bf k}_f\vert \varepsilon _i}                                  
L_{\mu \nu}^{(el)}W^{\mu \nu (el)}d^3{\bf k}_f,
\end{equation}   
\begin{equation}
d\sigma^{cc}=\frac{G^2 \cos^2\theta _c}{2}\frac{1}
{\varepsilon_i \varepsilon _f}                                         
L_{\mu \nu}^{(cc)}W^{\mu \nu (cc)}\frac{d^3{\bf k}_f}{(2 \pi)^2},
\end{equation}   
where $\alpha\simeq 1/137$ is the fine-structure constant, 
$G \simeq$ 1.16639 $\times 10^{-11}$ MeV$^{-2}$ is the Fermi constant, 
$\theta_c$ is the Cabbibo angle ($\cos \theta_c \approx$0.9749), 
$Q^2=-q^2=(k_i-k_f)^2$, and $q=(\omega,{\bf q})$ is the four-momentum 
transfer.

The lepton tensor can be written, by separating the symmetrical 
$l^{\mu \nu}_S$ and antisymmetric components $l^{\mu \nu}_A$ as follows
\begin{equation}
L^{\mu \nu (el)}=l^{\mu \nu}_S, ~~~~L^{\mu \nu (cc)}=l^{\mu \nu}_S +
l^{\mu \nu}_A,                                                          
\end{equation}   
\begin{equation}
l^{\mu \nu }_S=2(k^{\mu}_i k^{\nu}_f + k^{\mu}_f k^{\nu}_i -
g^{\mu \nu} k_i\cdot k_f),
\end {equation}                                                         
\begin{equation}
l^{\mu \nu }_A=-2 i \epsilon^{\mu \nu \alpha \beta}k_{i\alpha}k_{f\beta}, 
\end{equation}   
where $\epsilon^{\mu \nu \alpha \beta}$ is the antisymmetric tensor with 
$\epsilon^{0 1 2 3}=-\epsilon_{0 1 2 3}=1$.
Assuming the reference frame in which the {\it z}-axis is parallel to the 
momentum transfer ${\bf q}={\bf k}_i-{\bf k}_f$ and the {\it y}-axis is 
parallel to ${\bf k}_i \times {\bf k}_f$, the symmetrical components 
$l^{0x}_S, l^{x y}_S, l^{z y}_S$ and the anti-symmetrical ones $l^{0 x}_A, 
l^{x z}_A, l^{0 z}_A$, as well as those obtained from them by exchanging 
their indices, vanish.    

The electromagnetic $W^{(el)}_{\mu \nu}$ and weak charged-current 
$W^{(cc)}_{\mu \nu}$ hadronic tensors are given by bilinear products of the 
transition matrix elements of the nuclear electromagnetic (weak charged 
current) operator $J_{\mu}^{(el)(cc)}$ between the initial nucleus state 
$\vert A \rangle $ of energy $E_0$ and final state $\vert B_f \rangle$ of 
energy $E_f$ as
\begin{eqnarray}
W_{\mu \nu }^{(el)(cc)}&=&\sum_f \langle B_f\vert J^{(el)(cc)}_{\mu}
\vert A\rangle \times \langle A\vert J^{(el)(cc)\dagger}\vert 
B_f\rangle \nonumber\\ 
& & \times\delta (E_0+\omega -E_f),                                  
\end{eqnarray}   
where the sum is taken over the undetected states.

The transition matrix elements are calculated in the first order perturbation 
theory and in impulse approximation, {\it i.e.} assuming that the incident 
lepton interacts 
with the single nucleon while other ones behave as spectators. The nuclear 
current operator $J_{\mu}^{(el)(cc)}{\bf(q)}$ is taken as the sum of 
single-nucleon currents $j_{ \mu }^{(el)(cc)}({\bf q})$, {\it i.e.}  
$$
J_{\mu}^{\mu (el)(cc)}=\sum_{i=1}^{A}j_i^{\mu (el)(cc)},                            
$$
with
\begin{equation}
j_{\mu}^{(el)}=F_V(Q^2)\gamma _{\mu} + \frac{i}{2M}F_M(Q^2) 
\sigma_{\mu \nu}q^{\nu},                                                   
\end{equation}   
\begin{eqnarray}
j_{\mu}^{(cc)}&=&F_V(Q^2)\gamma _{\mu} + \frac{i}{2M}F_M(Q^2) 
\sigma_{\mu \nu}q^{\nu}\nonumber \\
& & + F_A(Q^2)\gamma_{\mu}\gamma ^5+ F_P(Q^2)q_{\mu}\gamma^5,              
\end{eqnarray}
where $M$ is the nucleon mass and $\sigma_{\mu \nu}=i[\gamma_{\mu}
\gamma_{\nu}]/2$. $F_V$ and $F_M$ are the isovector Dirac and Pauli nucleon 
form factors, taken from Ref.[6]. $F_A$ and $F_P$ are axial and 
pseudo-scalar form factors, parametrized as
\begin{equation}
F_A(Q^2)=\frac{F_A(0)}{(1+Q^2/M_A^2)^2}, ~~~                      
F_P(Q^2)=\frac{2M F_A(Q^2)}{m_{\pi}^2+Q^2},                        
\end{equation}   
where $F_A(0)=1.267$, $m_{\pi}$ is pion mass, and $M_A\simeq 1.032$ GeV is 
axial mass.

The general covariant form of the nuclear tensors is obtained in terms of 
 two four-vectors, namely the four-momenta of target $p^{\mu}$ and $q^{\mu}$.
 The electromagnetic and charged-current nuclear tensors can be written as
\begin{eqnarray}
W^{\mu \nu (el)}&=&-W_1^{(el)}g^{\mu \nu}+
\frac{W_2^{(el)}}{m_A^2}q^{\mu}q^{\nu}+
\frac{W_3^{(el)}}{m_A^2}p^{\mu}p^{\nu} \nonumber \\
& &+\frac{W_4^{(el)}}{m_A^2}(p^{\mu}q^{\nu}+p^{\nu}q^{\mu}),    
\end{eqnarray}
\begin{eqnarray}
W^{\mu \nu (cc)} & = & -W_1^{(cc)}g^{\mu \nu}+
\frac{W_2^{(cc)}}{m_A^2}q^{\mu}q^{\nu}\nonumber \\
& &+\frac{W_3^{(cc)}}{m_A^2}p^{\mu}p^{\nu} 
+\frac{W_4^{(cc)}}{m_A^2}(p^{\mu}q^{\nu}+p^{\nu}q^{\mu}) \nonumber \\  
& & + i\frac{W_5}{m_A^2}\epsilon^{\mu \nu \alpha \beta}q_{\alpha}p_{\beta},
\end{eqnarray} 
where $W_i$ are nuclear structure functions which depend on two scalars $Q^2$ 
and $p\cdot q$. 
Form Eqs.(10),(11) it follows that in our reference frame 
$W^{0x}, W^{0y}$, and $W^{xz}$, vanish together with tensor 
components obtained from them by exchanging their indices.
Therefore, obtained from contraction between lepton Eqs.(3),(4),(5) and 
nuclear Eqs.(10),(11) tensors, the inclusive cross sections for the QE 
electron (neutrino)-nucleus scattering are given by
\begin{equation}
\frac{d\sigma^{(el)}}{d\varepsilon_f d\Omega}=\sigma_M(v_L^{(el)} R_L^{(el)}+
v_T^{(el)} R_T^{(el)}),                                      
\end{equation}   
\begin{eqnarray}
\frac{d\sigma^{(cc)}}{d\varepsilon_f d\Omega}&=&\frac{G^2 \cos^2\theta_c}
{(2\pi)^2}
k_f \varepsilon_f (v_0 R^{(cc)}_L + v_T R^{(cc)}_T \nonumber \\
& & - v_{0L} R^{(cc)}_{0L}
+v_{LL} R^{(cc)}_{LL} \pm v^{(cc)}_{xy} R^{(cc)}_{xy}),        
\end{eqnarray}   
\begin{equation}
\sigma_M=\frac{\alpha ^2 \cos^2(\theta /2)}{4 \varepsilon^2_i 
\sin^4(\theta /2)},                                                   
\end{equation}   
where $\sigma_M$ is the Mott cross section and $\theta$ is the lepton 
scattering angle.  
The coefficients $v^{(el)}$ and $v$ are obtained from lepton tensors and 
can be written as  
\begin{figure*}
  \begin{center}
    \includegraphics[height=18pc,width=40pc]{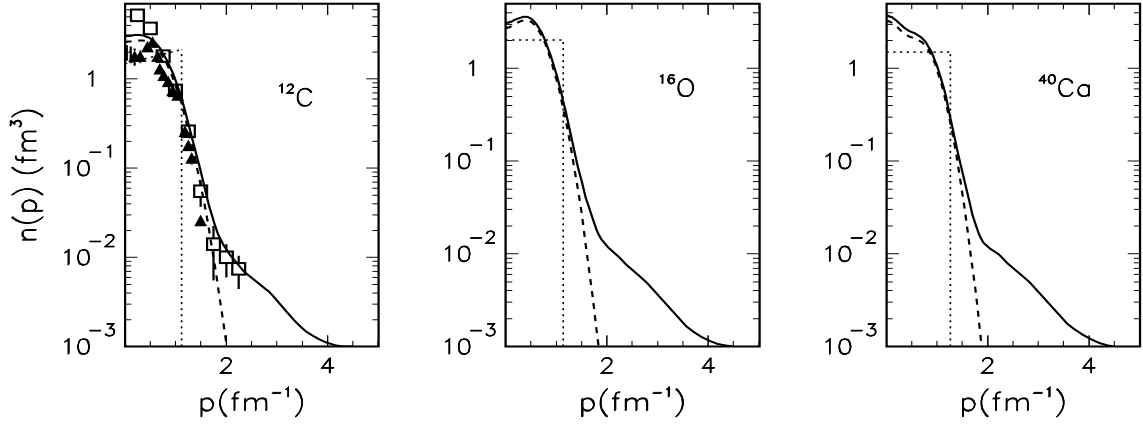}
  \end{center}
  \caption{Nucleon momentum distribution corresponding to Eq.(28) (solid lines)
 and Eq.(26) (dotted lines). The momentum distribution $n_0$ is given by 
short-dashed line. The open squares represent results obtained in Ref.[8]. 
The full triangles represent the values of $n_0(p)$ obtained in Ref.[9].  }
\end{figure*}
\begin{figure*}
\begin{center}
\includegraphics[height=40pc,width=40pc]{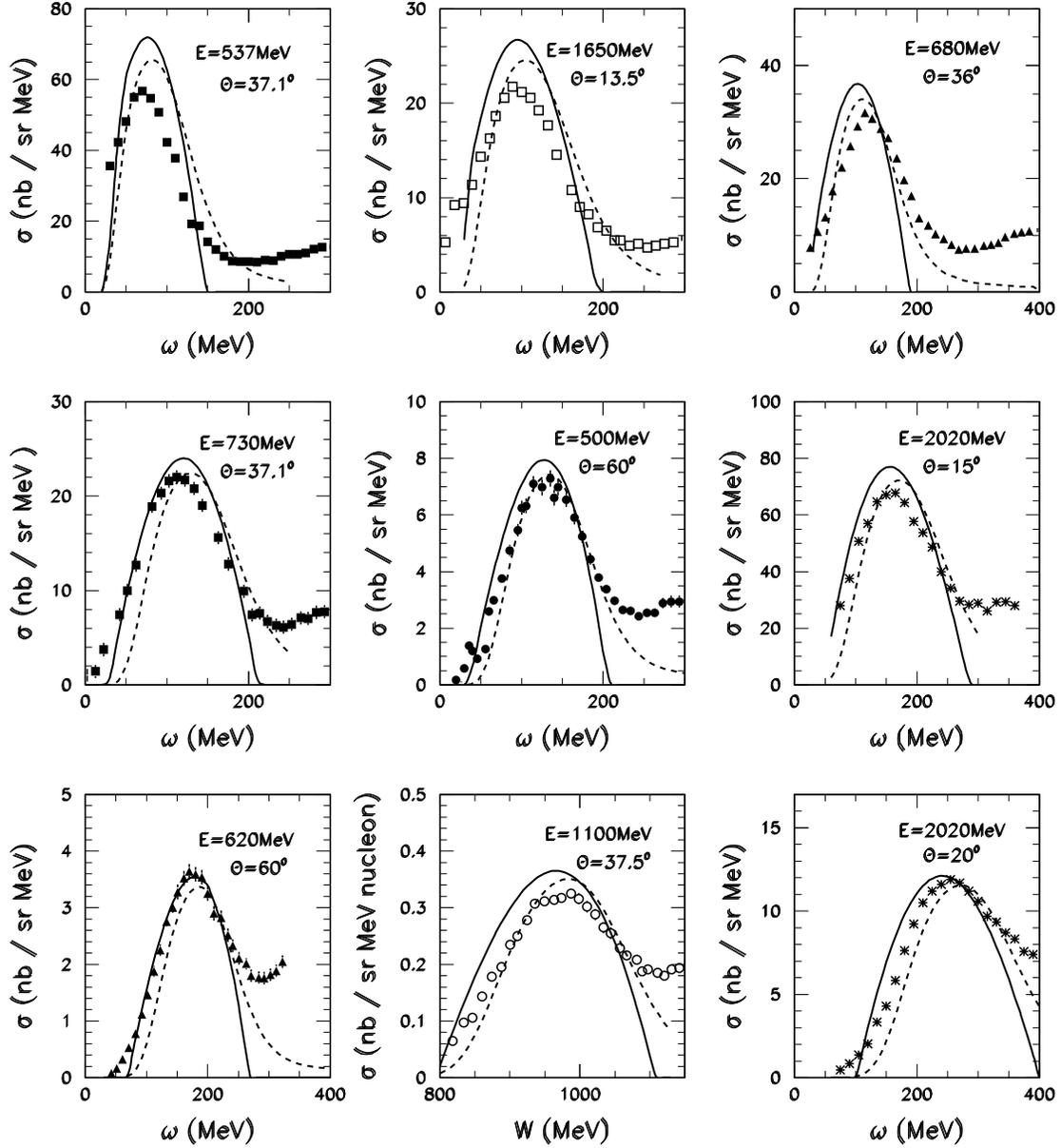}
\end{center}
\caption{Comparison of theoretical and experimental cross sections for 
$^{12}$C. The data are taken from Refs.[10] (filled circles), [11] 
(filled squares), [12] (filled triangles), [13] (open circles), [14] (open 
squares), and [15] (stars).}
\end{figure*}
\begin{figure*}
  \begin{center}
    \includegraphics[height=27pc,width=40pc]{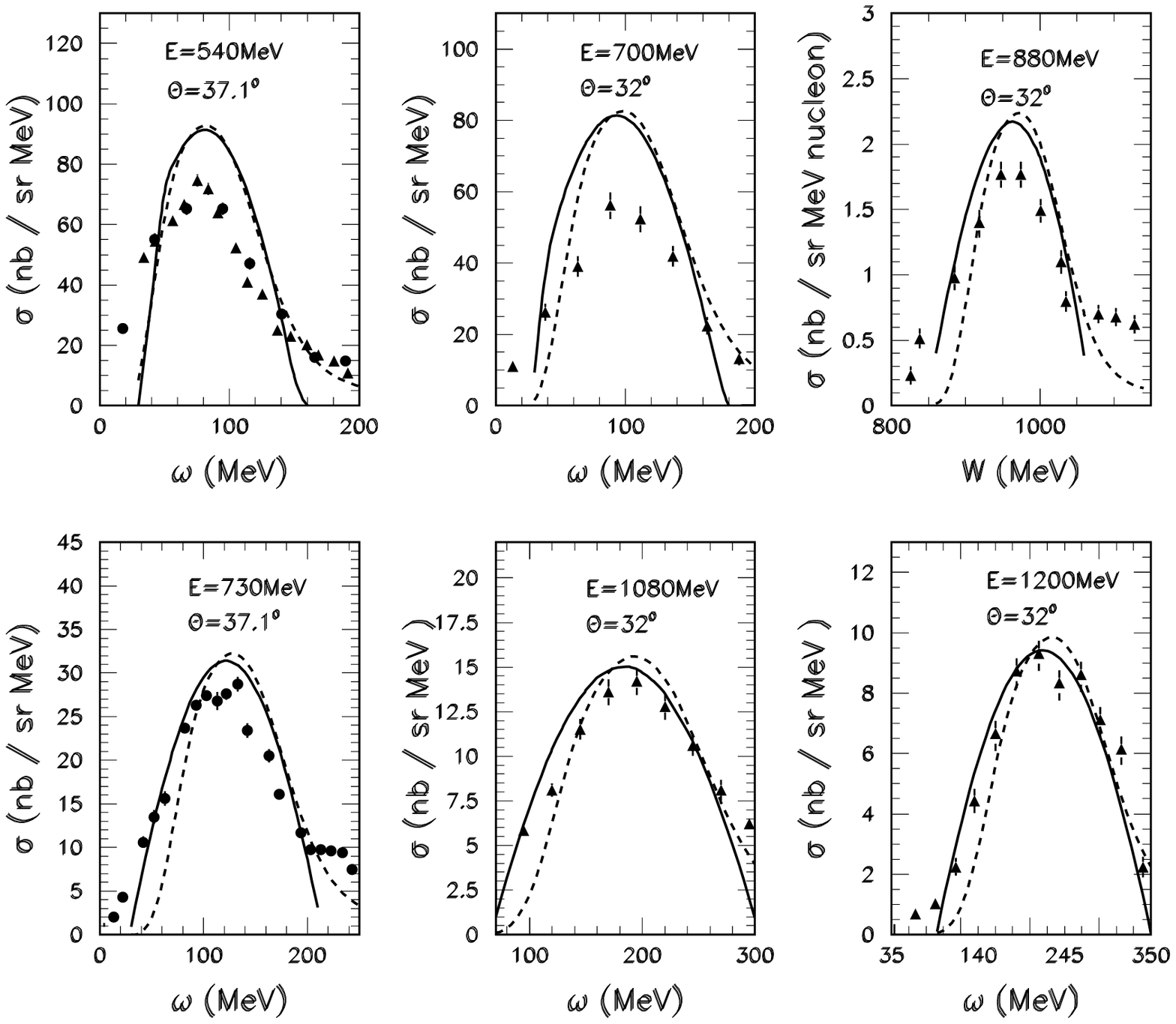}
  \end{center}
  \caption{Comparison of theoretical and experimental cross sections for 
$^{16}$O. The data are taken from Refs.[11] (filled circles), and [16] 
(filled triangles).}
\end{figure*}
\begin{figure*}
  \begin{center}
    \includegraphics[height=40pc,width=40pc]{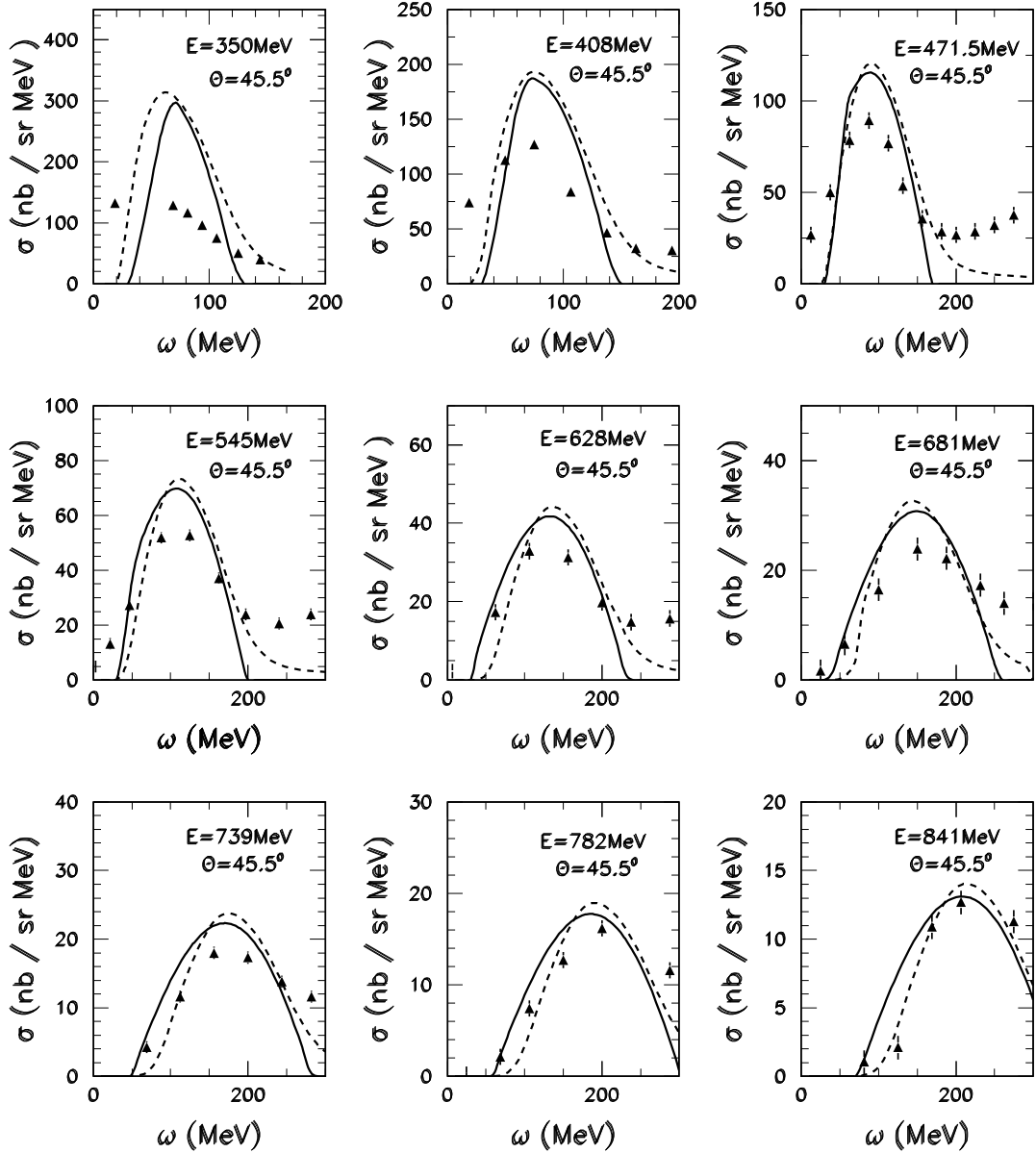}
  \end{center}
  \caption{Comparison of theoretical and experimental cross sections for 
$^{40}$Ca. The data are taken from Ref.[18].}
\end{figure*}
\begin{figure*}
  \begin{center}
    \includegraphics[height=27pc,width=40pc]{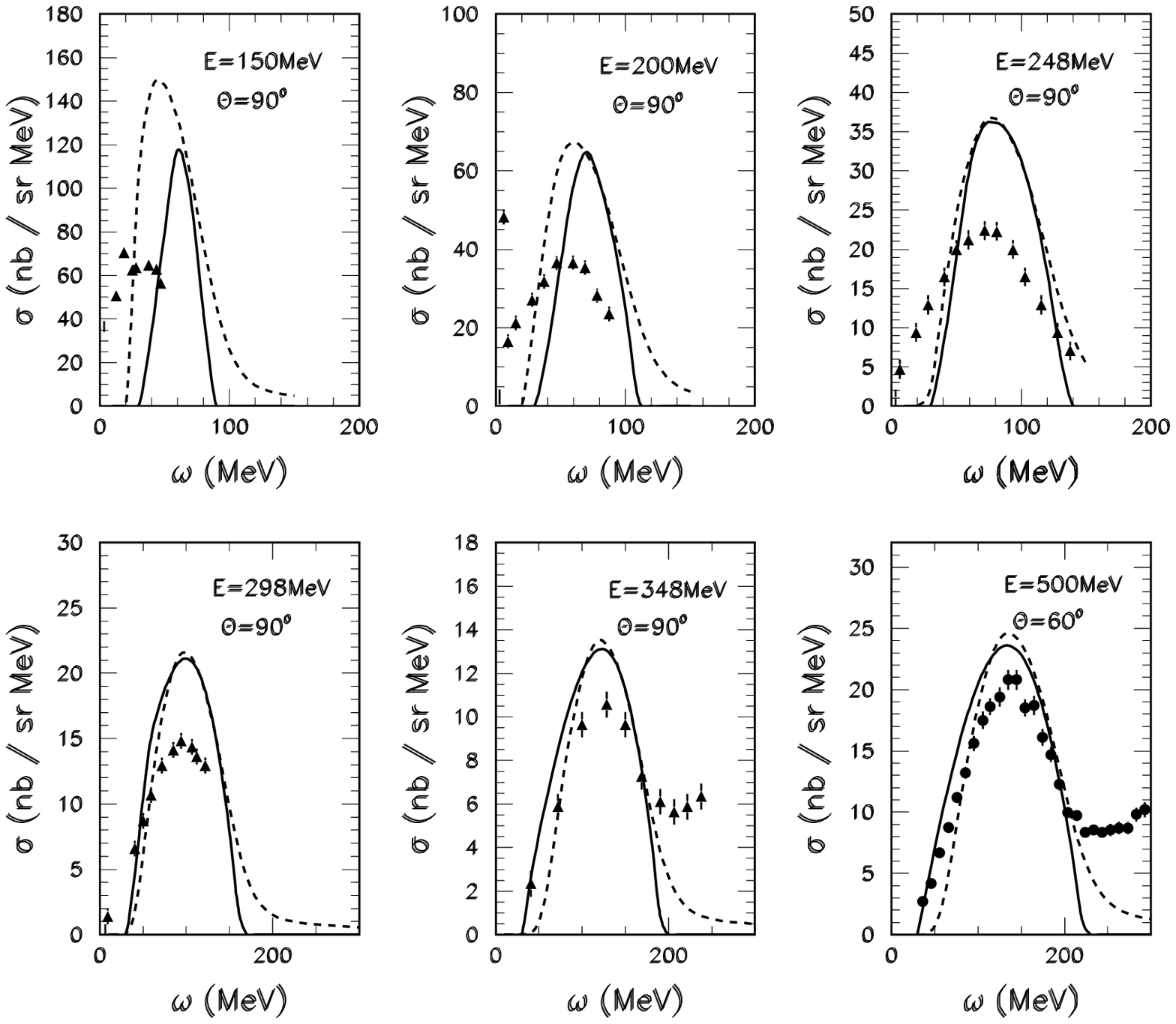}
  \end{center}
  \caption{Comparison of theoretical and experimental cross sections for 
$^{40}$Ca. The data are taken from Refs.[10] (filled circles), and [18] 
(filled triangles).}
\end{figure*}
\begin{figure*}
  \begin{center}
    \includegraphics[height=27pc,width=40pc]{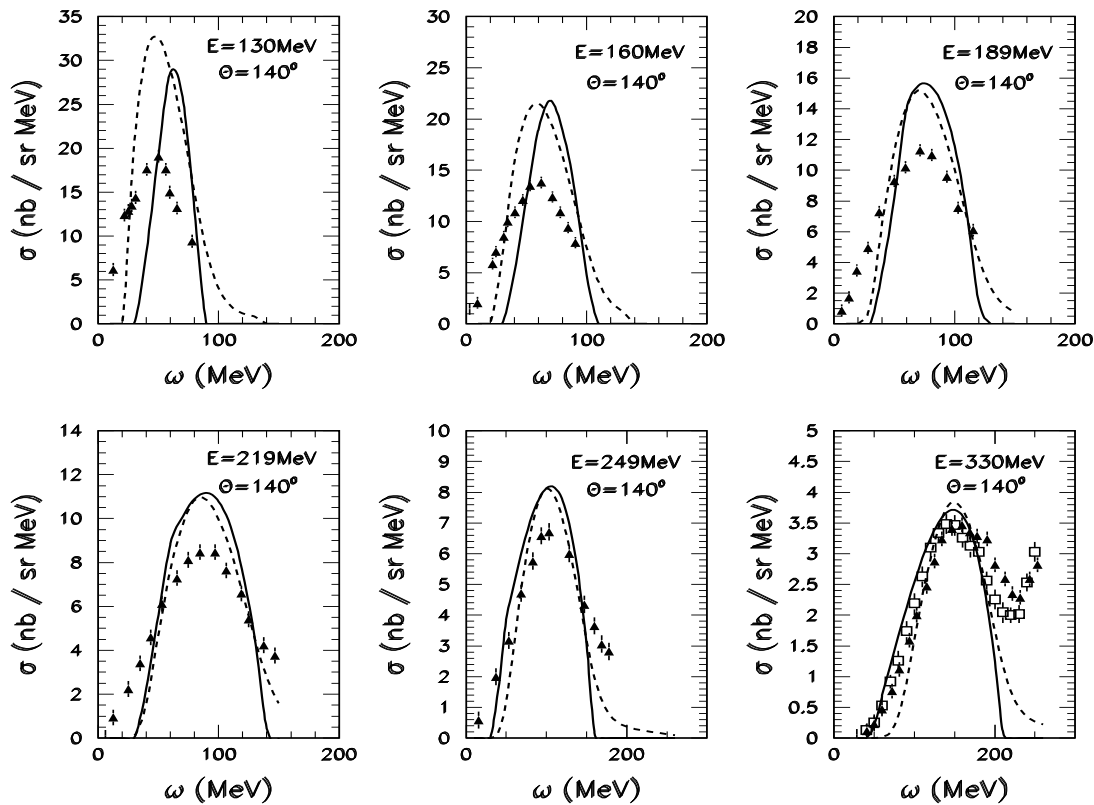}
  \end{center}
  \caption{Comparison of theoretical and experimental cross sections for 
$^{40}$Ca. The data are taken from Refs.[17] (open squares), 
and [18] (filled triangles).}
\end{figure*}
\begin{figure}
  \begin{center}
    \includegraphics[height=30pc,width=22pc]{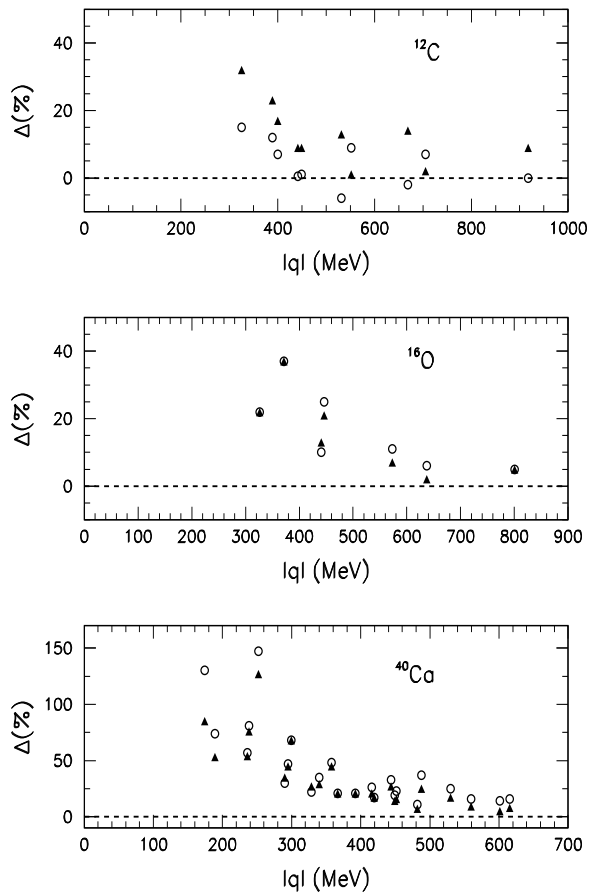}
  \end{center}
  \caption{ Differences between calculated and measured values of 
cross sections at maximum for $^{12}$C, $^{16}$O, and $^{40}$Ca as functions 
of three-momentum transfer $\vert {\bf q}\vert$. The filled triangles 
correspond to the Fermi gas model results and open circles correspond to the 
PWIA approach.}
\end{figure}
\begin{figure}[t]
  \begin{center}
   \includegraphics[height=18pc, width=22pc ]{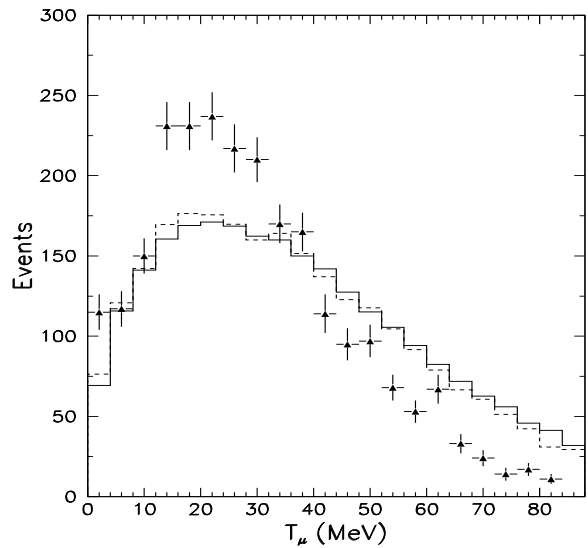}
  \end{center}
  \caption{ The distribution of muon kinetic energy for inclusive $^{12}$C
($\nu_{\mu},\mu^-$) reaction. Experimental data from Ref.[20]. The results of 
the RFGM (solid linen histogram) and the PWIA approach (short-dashed line 
histogram) are normalized to the data.}
\end{figure}
\begin{eqnarray}
v_L^{(el)}  =  \frac{Q^4}{\vert {\bf q}\vert ^4}, ~~~~~~ v_T^{(el)}=\frac{Q^4}
{2 \vert {\bf q}\vert ^2} + \tan^2(\theta /2),                         
\end{eqnarray}   
\begin{eqnarray}
 v_0 & = & 1+\hat{k} \cos \theta,\nonumber \\ 
v_T & = & 1-\hat{k}\cos \theta + 
\frac{\varepsilon_i\hat{k}\vert {\bf k}_f\vert}{\vert {\bf q}\vert^2}
\sin^2 \theta, \nonumber \\
v_{0L}&  = & \frac{\omega}{\vert{\bf q}\vert}(1 + \hat {k}\cos \theta)+
\frac{m^2_l}
{\vert {\bf q}\vert \varepsilon _f},\nonumber ~~\\
 v_{LL} & = & 1 + \hat{k}\cos \theta -2 
\frac
{\varepsilon _i \vert {\bf k}_f \vert \hat {k}}{\vert {\bf q} \vert ^2}
\sin^2 \theta, \nonumber \\ 
v_{xy} & = &\frac{\varepsilon_i+\varepsilon_f}{\vert {\bf q}\vert }(1-\hat{k} 
\cos\theta) -\frac{m^2_l}{\vert {\bf q}\vert \varepsilon_f},              
\end{eqnarray}   
where $\hat {k}=\vert{\bf k}_f\vert/\varepsilon_f$. 
The nuclear response functions $R^{(el)}$ and $R^{(cc)}$ 
are given in terms of components of nuclear tensors as follows [4],[7]
\begin{eqnarray}
R^{(el)}_L &=& W^{(el)}_{00}=\frac{\vert{\bf q}\vert^2}{Q^2}W_1^{(el)}+
W_2^{(el)},
\nonumber \\
R_T^{(el)} &=& W_{xx}^{(el)}+W_{yy}^{(el)}=-2W_1^{(el)}+W_3^{(el)},
\nonumber \\ 
R_L^{(cc)} &=& W^{(cc)}_{00}=-W_1^{(cc)}+\frac{\omega ^2}{m_A^2}W_2^{(cc)}+
W_3^{(cc)} \\
 & & +\frac{2\omega}{m_A}W_4^{(cc)}, \nonumber \\ 
R_{0L}^{(cc)} &=& W^{(cc)}_{0z}+W^{(cc)}_{z0}=\frac{2\vert{\bf q}\vert}{m_A^2}
(\omega W_2^{(cc)}+ m_A W_4^{(cc)}),\nonumber  \\
R_T^{(cc)} &=& W_{xx}^{(cc)}+W_{yy}^{(cc)}=2W_1^{(cc)}, \nonumber \\ 
R_{LL}^{(cc)} &=& W^{(cc)}_{zz}=W_1^{(cc)} + \frac{\vert {\bf q}\vert^2}{m_A^2}
W_2^{(cc)}, \nonumber \\  
R^{(cc)}_{xy} &=& \frac{2\vert{\bf q}\vert}{m_A}W^{(cc)}_5.                 
\end{eqnarray}
In order to evaluate nuclear response functions we consider the RFGM  
and PWIA approach based on assumption 
that the virtual photon interacts with off-shell nucleon and 
neglecting interaction of the knocked out nucleon with the residual nucleus. 
In the PWIA the four-momenta of the 
initial nucleus A, the bound off-shell nucleon N, and the final state B are
\begin{eqnarray}
& p & \equiv (m_A,{\bf 0}),~~~~~~ p\equiv (m_A-({\bf p}^2+m^{\ast 2}_B)^{1/2},
{\bf p}), \nonumber  \\ 
& p_B & \equiv  (({\bf p}^2+m^{\ast 2}_B)^{1/2},-{\bf p}),
\nonumber
\end{eqnarray}
respectively. Here $m^{\ast}_B=m_B+\epsilon _f$, $m^{\ast}_B$ and 
$\epsilon_f$ are 
the mass and intrinsic energy of the final (A-1)-nucleon state, respectively.
Within the above assumption the nuclear structure functions can be written  
as follows
\begin{eqnarray}
W^{A(el)}_i &=&\int d{\bf p} \int d E Z S^p(\vert {\bf p}\vert ,E)        
\sum_{j=1}^2C_{i j}W_j^{p,off}(Q^2) \nonumber \\
& & +\textrm{similar terms for the neutrons,}
\end{eqnarray}   
\begin{eqnarray}   
W^{A(cc)}_i &=& \int d{\bf p} \int d E (A-Z) S^n(\vert {\bf p}\vert ,E)
\nonumber \\     
& & \times \sum_{j=1}^5D_{i j}W_j^{n,off}(Q^2).
\end{eqnarray}   
Here, $Z$ is the number of protons, $W_j^{N,off} (N=p,n)$ are the off-shall 
nucleon structure functions that are given in the terms of nucleon 
form-factors. $S^N(\vert{\bf p}\vert,E)$ is the nucleon spectral function and 
kinematic factors $C_{i j}$, $D_{i j}$ have form
\begin{eqnarray}   
C_{1 1} = 1, ~~~~~~ C_{1 2}=\left[ \vert {\bf p} \vert^2 -\left
(\frac{{\bf p q}}{\vert{\bf q}\vert}\right)^2 \frac{1}{2M^2}\right],~~~~~~
C_{2 1}=0, \nonumber 
\end{eqnarray}
\begin{equation}
C_{2 2}= \frac{1}{M^2}\left[ p_0^2-2p_0 \omega \frac{{\bf p q}}
{\vert {\bf q} \vert ^2} + \frac{\omega^2}{\vert {\bf q}\vert^2}\left( 
\frac{{\bf p q}}{\vert {\bf q}\vert}\right)^2 \right] 
-\frac{Q^2}{\vert{\bf q}
\vert^2}C_{1 2},         
\end{equation}   
and
\begin{eqnarray}   
D_{1 1}=1,~D_{1 3}=\frac{\vert{\bf p}\vert^2}{2M^2}(1-\cos^2\tau),\nonumber
\end{eqnarray}   
\begin{eqnarray}   
D_{1 2}=D_{1 4}=D_{1 5}=0,\nonumber 
\end{eqnarray}   
\begin{eqnarray}   
D_{2 2} &=& \frac{m_A^2}{M^2}, ~~~~ 
D_{2 3}=\frac{\vert{\bf p}\vert^2m^2_A(3\cos^2\tau-1)}
{2M^2{\vert \bf q}\vert^2},\nonumber \\
D_{2 4} &=& 2\left(\frac{m_A}{M}\right)^2\frac{\vert
{\bf p}\vert}{\vert{\bf q}\vert}\cos \tau,~~~~D_{2 1}=D_{2 5}=0,\nonumber
\end{eqnarray}   
\begin{eqnarray}   
\lefteqn{ D_{3 3}= \frac{\vert{\bf p}\vert^2}{2M^2}
\left(1-\frac{\omega^2}{\vert {\bf q}\vert^2}\right)(1-\cos^2\tau) {} }
                                                        \nonumber \\
& & {}+\frac{\vert{\bf p}\vert^2\omega^2\cos^2\tau}{M^2\vert {\bf q}\vert^2}+
\frac{2p_0\omega}{M^2} \frac{\vert{\bf p}\vert}{\vert {\bf q}\vert}\cos\tau 
+\frac{p_0^2}{M^2} ,\nonumber
\end{eqnarray}   
\begin{eqnarray}  
D_{3 1}=D_{3 2}=D_{3 4}=D_{3 5}=0,\nonumber 
\end{eqnarray}   
\begin{eqnarray}   
D_{4 3} &=& -\frac{m_A}{M^2}\left [\frac{p_0\vert {\bf p}\vert}
{\vert {\bf q}\vert}\cos\tau - \frac{\omega\vert{\bf p}\vert^2}
{2\vert{\bf q}\vert^2}(3\cos^2\tau-1)\right], \nonumber \\
D_{4 4} &=& \frac{m_A}{M^2}\left(p_0+\frac{\omega \vert{\bf p}\vert}
{\vert {\bf q}\vert}\cos\tau \right),~D_{4 1}=D_{4 2}=D_{4 5}=0,\nonumber
\end{eqnarray}   
\begin{eqnarray}   
D_{5 1} &=& D_{5 2}=D_{5 3}=D_{5 4}=0,\nonumber \\
D_{5 5} &=& \frac{m_A}{M^2}\left( p_0 +
\frac{\omega}{\vert {\bf q}\vert} \vert{\bf p}\vert \cos\tau\right),     
\end{eqnarray}   
where $\cos \tau ={\bf p}\cdot {\bf q}/\vert {\bf p}\cdot {\bf q}\vert$.
In this paper 
we assume that $W_j^{N,off}$ are identical to free nucleon structure 
function $W_j^N$. The parametrization of $W_j^N$ is taken from Refs.[2,3] as 
follows
\begin{equation}
W_1^{N (el)}=\frac{Q^2}{4p_0p'_0}[F_V(Q^2)+F_M(Q^2)]^2
\delta(p_0+\omega-p'_0),                                               
\end{equation}
\begin{eqnarray}
W_2^{N (el)} &=& \frac{M^2}{p_0p'_0}[F_V^2(Q^2)+
\frac{Q^2}{4M^2}F_M^2(Q^2)]
\nonumber \\
& & \times \delta(p_0+\omega-p'_0),                                       
\end{eqnarray} 
and  
\begin{equation}
W_j^{n,(cc)}=\frac{1}{2p_0p'_0}\widetilde{W}_j\delta(p_0+\omega-p'_0),    
\end{equation}
with
\begin{eqnarray}   
\widetilde{W}_1&=&\frac{Q^2}{2}\left[(F_V+F_M)^2+F_A^2\left(1+\frac{4M^2}
{Q^2}\right)\right],\nonumber \\
\widetilde{W}_3&=& 2M^2\left [F_V^2+\frac{Q^2}{4M^2}F_M^2+F_A^2\right],
\nonumber \\
\widetilde{W}_2&=&\frac{1}{4}\widetilde{W}_3-\frac{M^2}{Q^2}\widetilde{W}_1+
\frac{2}{Q^2}\left(F_A+\frac{Q^2}{2M}F_P\right)^2, \nonumber \\
\widetilde{W}_4&=&\frac{1}{2}\widetilde{W}_3,~~~~~~                
\widetilde{W}_5=2M^2F_A(F_V+F_M),
\end{eqnarray}   
where $p'_0=[({\bf p}+{\bf q})^2+M^2]^{1/2}$ is energy of the knocked-out 
nucleon.

The nucleon spectral function $S^N(\vert {\bf p}\vert,E)$ in the PWIA 
represents probability to find the nucleon with the momentum ${\bf p}$ 
and the removal energy $E$ in the ground state of the nucleus. 
In the commonly used Fermi gas model, that was described by 
Smith and Moniz [2], nucleons in nuclei are assumed to occupy plane wave 
states in uniform potential while the knocked-out nucleon  
is outside of the Fermi sea. The Fermi gas model provides the simplest form 
of the spectral function which is given by
\begin{eqnarray}
S_{FG}(E,\vert {\bf p}\vert)&=&\frac{3}{4\pi p_F^3}
\Theta(p_F-\vert{\bf p}\vert) 
\Theta(\vert{\bf p} + {\bf q}\vert-p_F) \nonumber \\
& & \times \delta[(\vert{\bf p}\vert^2+M^2)^{1/2}-\epsilon_b-E)],      
\end{eqnarray} 
where $p_F$ denotes the Fermi momentum and a parameter $\epsilon_b$ is 
effective binding energy, introduced to account of nuclear binding.
The QE lepton-nucleus reactions are complicated processes, involving nuclear 
many body effects. The calculation of the nuclear spectral function for 
complex nuclei requires to solve many body problem. In this paper we consider 
also a phenomenological model using PWIA approach with the spectral function 
which 
incorporates both the single particle nature of the nucleon spectrum at low 
energy and high-energy and high momentum components due to NN-correlations 
in ground state. Following [4,5] we separate the full spectral function into 
two parts  
\begin{equation}
S(E,{\bf p})=S_0(E,{\bf p}) + S_1(E,{\bf p}).                    
\end{equation} 
The integration of Eq.(27) over energy gives nucleon momentum distribution,
\begin{equation}
n({\bf p})=\int \frac{dE}{2\pi}S(E,{\bf p})=n_0({\bf p})+n_1({\bf p}).   
\end{equation} 
The spectral function is normalized according to
\begin{equation}
\int \frac{dE d{\bf p}}{2\pi}S(E,{\bf p})=1.                             
\end{equation} 
The low-energy part ($S_0$) can be approximated by the following expression
\begin{equation}
S_0(E,{\bf p})=2\pi n_0({\bf p})\delta(E- \varepsilon^{(1)} -
{\bf p}^2/2M_{A-1}),                                                    
\end{equation} 
with $\varepsilon^{(1)}$ the nucleon separation energy averaged over residual 
configurations of $A-1$ nucleons with low excitation energies and recoil 
energy ${\bf p}^2/2M_{A-1}$, $n_0(\vert {\bf p}\vert)$ is the 
corresponding part of nucleon momentum distribution.

The high-excitation part $(S_1)$ of the spectral function is determined by 
excited states with one or more nucleons in continuum.
The detailed 
description of this model is given in Refs.[4,5] as well as parametrization 
of $n_0({\bf p})$ and 
$n_1({\bf p})$, which fit the result of many-body calculations of nuclear 
momentum distribution. As follows from these 
calculations the low momentum part incorporates about 80\% of the total 
normalization of spectral function, while the other 20\% are taken by the high 
momentum part.
The nucleon momentum distributions $n(\vert {\bf p}\vert)$ and $n_{FG}(\vert 
{\bf p}\vert)$ are shown in Fig.1. The normalization of $n(p)$ and $n_{FG}(p)$ 
is $\int dp p^2 n(p)=1$, where $p=\vert {\bf p} \vert$. The distributions 
$n_{FG}$ for various nucleus $^{12}$C, $^{16}$O and $^{40}$Ca were calculated 
using the value of parameters $p_F=221$ MeV, $\epsilon_b=25$ MeV ($^{12}$C); 
$p_F=225$ MeV, $\epsilon_b=27$ MeV ($^{16}$O); and $p_F=249$ MeV, 
$\epsilon_b=33$ MeV) ($^{40}$Ca) [10].  

\section{Results}

There is vast high-precision data for electron scattering off 
nucleus $^{12}$C, $^{16}$O, and $^{40}$ Ca. Hence these nuclei are 
taken at the focus of the present work.
Data on inclusive cross sections for a number of nuclei 
(A between 6 and 208) but only one set of kinematics were obtained early in 
Ref.[10]. Carbon data are available from experiments 
[11]-[15]. For oxygen target the experiments were performed by SLAC [11] and 
Frascaty [16] groups. For calcium target the inclusive cross section have been 
measured in experiments [10], 
[17]-[19]. All these data were used in our analysis.    
   
Using both the relativistic Fermi gas model and the PWIA approach described 
above, we calculated the inclusive cross sections for given kinematics 
(energies and angles) and compared them with data. The results are 
presented in Figs.2-6 for $^{12}$C, $^{16}$O, and $^{40}$Ca respectively. 
The solid lines 
are the results in the Fermi gas model, while short-dashed 
lines are results in the PWIA. The differences can be 
seen from these figures in which the cross sections as functions 
of $\omega$ or invariant mass produced on a free nucleon $W$ are plotted. 
At the maximum of the cross sections both models overestimate the measured 
values. We evaluated the differences between predicted ($\sigma_{cal}$) 
and measured ($\sigma_{data}$) quantities 
$\Delta=\sigma_{calc}-\sigma_{data}$. $\Delta(\vert {\bf q}\vert)$ 
as a function of three-momentum transfer $\vert {\bf q}\vert$, is shown 
in Fig.7, from which it is clear that the $\Delta(\vert {\bf q} \vert)$ 
decreases with $\vert {\bf q}\vert$ from 30$\div $50\% at 
$\vert {\bf q}\vert\leq$ 
200 MeV to 10$\div $15\% at $\vert {\bf q}\vert \geq$ 500 MeV. 

In Refs.~\cite{REV17},~\cite{REV18} transverse $R_T^{(el)}$ and longitudinal 
$R_L^{(el)}$ 
response functions have been extracted for 200 MeV$\leq\vert {\bf q}\vert\leq$
500 MeV . It has been shown that the RFGM 
overestimates 
the observed longitudinal response for about 40\% ~\cite{REV17} 
($\sim $20\% ~\cite{REV18}). 
At low $\vert {\bf q}\vert$ this model also overestimates the magnitude of the 
transverse response function. At higher values of $\vert {\bf q}\vert$ it 
do better at reproducing the value of the $R_T^{(el)}$. 

The predictions of both models are compared with the experimental result 
of the LSND collaboration at Los Alamos for $^{12}$C$(\nu_{\mu},\mu^-)$ 
reaction~\cite{REV20}. The calculations are flux-averaged over the Los 
Alamos neutrino flux. The mean energy of neutrino flux above threshold is 
156 MeV. The comparison is shown 
in Fig.8 where the calculated muon energy distributions are normalized to the 
experimental total number events. We note that both models do 
not give an accurate 
descriptions of the shape of the muon spectrum. The flux-averaged cross 
section integrated over the muon energy is 17.8$\times$10$^{-40}$ cm$^2$ 
in the case of the RFGM and 26.8$\times$10$^{-40}$ cm$^2$ in the PWIA. 
The experimental value is (10.6$\pm$0.3$\pm$1.8)$\times$10$^{-40}$ cm$^2$. 
The result obtained by other calculation in the framework of the Fermi gas 
model with local density approximation [21] gives also larger value 
$\sigma$=(16.7 $\pm$ 1.37)$\times$10$^{-40}$ cm$^2$.        
         
\section{Conclusions}

In this work we have tested the widely used relativistic Fermi gas model and 
plane-wave impulse approximation approach against electron-nucleus scattering
data. We calculated the inclusive QE cross sections and compared them with 
high-precision data for $^{12}$C, $^{16}$O, and $^{40}$Ca in a wide region of 
incident energy and momentum. We evaluated the differences 
$\Delta$ between predicted and measured QE cross section 
at the maximum and found that both models overestimate the measured values. 
The function $\Delta(\vert{\bf q}\vert)$ decreases with  
three-momentum transfer from 30$\div$50 \% at $\vert {\bf q}\vert \leq$ 
200 MeV to 10$\div$15 \% at $\vert {\bf q}\vert \geq$ 500 MeV. Therefore 
these models overestimate also the cross sections at low 
$Q^2=\vert {\bf q}\vert^2-\omega ^2$.  

We applied the RFGM and PWIA approach to $^{12}$C($\nu_{\mu},\mu^-$) reaction. 
The flux-averaged total cross sections and muon energy distributions were 
calculated and compared with experimental results of the LSND collaboration. 
The calculated cross sections are significantly larger than the 
experimental ones and both models do not give an accurate description of 
the shape of muon spectrum.

In conclusion we note that the inclusion of final state interaction effects 
along with realistic spectral function may significantly correct the 
description of the data at low momentum transfer, as was pointed in Ref.[16].

\begin{acknowledgments}
This work was supported by the Russian Foundation for Basic Research project 
No 02-02-17036. We would like to thank S. Kulagin for fruitful discussions. 
\end{acknowledgments}



\end{document}